\documentclass[aps,showpacs,superscriptaddress,twocolumn]{revtex4}

\usepackage{amsfonts}
\usepackage{amsmath}
\usepackage{amssymb}
\usepackage{graphicx}

\input{tcilatex}

\begin{document}
\title{First-principles study of \textbf{$\alpha$}-Pu$_{2}$O$_{3}$}
\author{Hongliang Shi}
\affiliation{Institute of Applied Physics and Computational
Mathematics, P.O. Box 8009, Beijing 100088, People's Republic of
China} \affiliation{SKLSM, Institute of Semiconductors, Chinese
Academy of Sciences, P. O. Box 912, Beijing 100083, People's
Republic of China}
\author{Ping Zhang}
\thanks{Author to whom correspondence should be addressed. Electronic
address: zhang\_ping@iapcm.ac.cn} \affiliation{Institute of Applied
Physics and Computational Mathematics, P.O. Box 8009, Beijing
100088, People's Republic of China} \affiliation{Center for Applied
Physics and Technology, Peking University, Beijing 100871, People's
Republic of China} 

\begin{abstract}
We systematically investigate the electronic structure, magnetic
order, and valence states of $\alpha$-Pu$_{2}$O$_{3}$
(\emph{C}-type) by using first-principles calculations.
$\alpha$-Pu$_{2}$O$_{3}$ can be constructed from PuO$_{2}$ by
removing 25\% oxygen atoms. Our results show that the Pu 5\emph{f}
orbitals are further localized after removing ordered oxygen atoms.
This phenomenon is demonstrated by the combined fact that (i) the
volume per unit cell expands 7\% and (ii) the corresponding magnetic
moments and valence states for Pu ions increase and decrease,
respectively. According to the density of states and charge density
distribution analysis, PuO$_{2}$ is found to be more covalent than
$\alpha$-Pu$_{2}$O$_{3}$, which is also because of the more
localization of 5\emph{f} orbitals in the latter. The calculated
lattice constants, bulk modulus, and electronic structures for
PuO$_{2}$ and $\alpha$-Pu$_{2}$O$_{3}$ are consistent well with
experimental observations.
\end{abstract}

\maketitle

\section{INTRODUCTION}

Plutonium-based materials have been extensively investigated because of not
only their great technological importance in the nuclear industry application
but also their interesting rich physical properties from the basic theoretical
viewpoint. Metallic Pu locates in the boundary of localized and delocalized
5\emph{f} electrons among the actinide metals and it has six different phase
under different temperatures and pressures because of the complex character of
the 5\emph{f} electrons \cite{r1,r2}. As for plutonium oxides, which are the
only products when metallic plutonium is exposed in air, can store the surplus
metallic plutonium \cite{r3}. The corrosion oxidation of metallic plutonium in
air is very hot topic since it is a key problem for the protection and storage
of plutonium-based nuclear weapons. Furthermore, the thermodynamics and redox
properties of plutonium oxides are also complex and interesting. When Pu is
exposed to dry air at room temperature, the plutonium dioxide PuO$_{2}$ layer
is formed and a thin layer of plutonium oxide Pu$_{2}$O$_{3}$ is followed on
plutonium surfaces \cite{r3}. After several months or years, most of the
PuO$_{2}$ layer auto-reduces into Pu$_{2}$O$_{3}$ layer. Especially, at
150-200$^{\mathrm{{o}}}$C, all the PuO$_{2}$ layer auto-reduces into Pu$_{2}%
$O$_{3}$ layer in minutes \cite{r3}. Usually, the mentioned plutonium
sesquioxide is $\beta$-Pu$_{2}$O$_{3}$ in the hexagonal structure
(\emph{P$\bar{3}$m\textrm{{1}}}). However, another abnormal body centered
cubic form, plutonium sesquioxide PuO$_{1.52}$, was also detected \cite{r4}.
The ideal stoichiometric cubic plutonium sesquioxide is $\alpha$-Pu$_{2}%
$O$_{3}$ (\emph{C}-type PuO$_{1.5}$), which are consisted of 32 Pu atoms and
48 O atoms per unit cell. $\alpha$-Pu$_{2}$O$_{3}$ has not been prepared as a
single-phase compound, since it is stable only below 300$^{\mathrm{{o}}}$C
\cite{r5}. A mixture of cubic PuO$_{1.52}$ and PuO$_{1.98}$ can be obtained by
partial reduction of PuO$_{2}$ at high temperature and cooling to room
temperature \cite{r5}.

Naturally, the reduction of PuO$_{2}$ into $\alpha$-Pu$_{2}$O$_{3}$ is easy to
happen because of the similarity of their cubic structures. PuO$_{2}$
crystallizes in the CaF$_{2}$ structure (see Fig. 1) with the plutonium and
oxygen atoms forming face-centered and simple cubic sublattices, respectively,
and the cubic $\alpha$-Pu$_{2}$O$_{3}$ can be obtained from PuO$_{2}$
2$\times$2$\times$2 supercell after removing ordered 25\% oxygen atoms
\cite{r3}. Actually, the structure of $\alpha$-Pu$_{2}$O$_{3}$ is
body-centered cubic with space group \emph{Ia$\bar{3}$} (No.206). Oxygen atoms
occupy the 48\emph{e} sites, and plutonium atoms occupy the 24\emph{d} and
8\emph{a} sites \cite{r5} (see Fig. 1). Furthermore, if the 16\emph{c} sites
are also occupied by oxygen atoms, the resulting cell is the same as PuO$_{2}$
2$\times$2$\times$2 supercell. Therefore, as mentioned above, the cubic
$\alpha$-Pu$_{2}$O$_{3}$ can be obtained by removing oxygen atoms located in
the 16\emph{c} (0.25,0.25,0.25) sites.

\begin{figure}[ptb]
\includegraphics*[height=4.6cm,keepaspectratio]{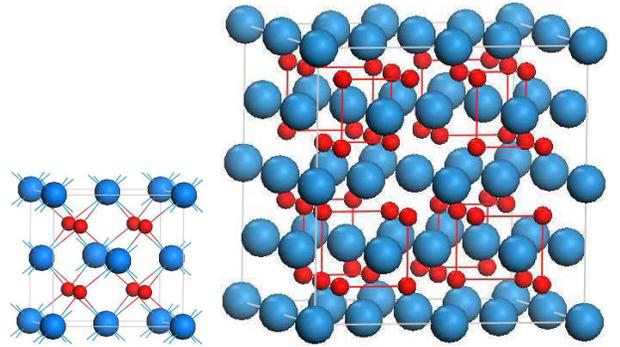}\caption{(left) unit cell
of PuO$_{2}$ with CaF$_{2}$-type structure. (right) unit cell of $\alpha
$-Pu$_{2}$O$_{3}$ (body-centered cubic). The blue and red spheres denote Pu
and O atoms, respectively.}%
\label{fig1:epsart}%
\end{figure}

In this work, we pay special attention to how the electronic structure,
magnetic order, and Pu valence states change during the PuO$_{2}$-Pu$_{2}%
$O$_{3}$($\alpha$) reduction process. In PuO$_{2}$, Pu is in the
ideal stable $+$4 oxidation state while in $\alpha$-Pu$_{2}$O$_{3}$
the ideal oxidation state is $+3$. This change of valence states may
illustrates that the 5\emph{f} electrons are more localized in
$\alpha$-Pu$_{2}$O$_{3}$. Similar to CeO$_{2}$-Ce$_{2}$O$_{3}$
reduction transition \cite{r6}, after creating one oxygen vacancy,
two electrons left behind are condensed into localized
\emph{f}-level traps on two plutonium atoms, therefore, the valence
for Pu changes from +4 to +3. Furthermore, due to the localization
our results show that the volume expands 7\% during
PuO$_{2}$-Pu$_{2}$O$_{3}$ isostructure reduction process.

In order to describe more accurately the strong on-site Coulomb
repulsion interaction among the plutonium 5\emph{f} electrons, we
use the generalized gradient approximation (GGA)+\emph{U} scheme.
Our previous studies showed that the GGA+\emph{U} approach can
accurately describe the electronic structures and thermodynamic
properties of PuO$_{2}$ and $\beta$-Pu$_{2}$O$_{3}$ \cite{r8}, which
has motivated more theoretical calculations in these two years
\cite{Jom,Andersson,Petit}. In the following, our calculated results
demonstrate that the GGA+\emph{U} correction can also successfully
predict the ground state properties of cubic
$\alpha$-Pu$_{2}$O$_{3}$ and reliably describe the
PuO$_{2}$-Pu$_{2}$O$_{3}$($\alpha$) reduction process. Our study
shows that the system's volume expands during the reduction process.
Also, two well-resolved peaks in the density of states (DOS) are
observed and identified to originate from the Pu 5$f$ and O 2$p$
states in PuO$_{2}$, which are consistent well with the recent
photoemission measurements \cite{r9}. In addition, our calculated
bulk modulus for PuO$_{2}$ is 180 GPa with Hubbard \emph{U}=3.0 eV,
which agrees excellently with the recent refined experimental values
of 178 GPa \cite{r90}. However, in Ref. \cite{r10}, this value is
largely overestimated to be 379 GPa.

\section{DETAILS OF CALCULATION}

Our first-principles calculations are based on the density
functional theory (DFT) and the Vienna ab initio simulation package
(VASP) \cite{r11} using the GGA for the exchange correlation
potential \cite{r12}. The electron and core interactions are
included using the frozen-core projected augmented wave (PAW)
approach which combines the accuracy of augmented-plane-wave methods
with the efficiency of the pseudo-potential approach\cite{r13}. The
Pu 5\emph{f}, 6\emph{s}, 6\emph{p}, 6\emph{d}, and 7\emph{s} as well
as the oxygen 2\emph{s} and 2\emph{p} electrons are explicitly
treated as valence electrons. The electron wave function is expanded
in plane waves up to a cutoff energy of 500 eV. For the Brillouin
zone integration, the 2$\times$2$\times$2 Monkhorst-Pack sampling is
adopted. The strong on-site Coulomb repulsion among the localized Pu
5\emph{f} electrons is described by using the formalism formulated
by Dudarev \emph{et al. }\cite{r7}. In this scheme, only the
difference between the spherically averaged screened Coulomb energy
\emph{U} and the exchange energy \emph{J} is important for the total
LDA (GGA) energy functional. Thus, in the following we label them as
one single effective parameter \emph{U} for brevity. In our
calculation, we use \emph{J}=0.75 eV for the exchange energy, which
is close to the values used in other previous work
\cite{r8,Andersson}.

\section{RESULTS AND DISCUSSIONS}

\subsection{Atomic and electronic structure of PuO$_{2}$}

In order to investigate the electronic and structural properties of
$\alpha $-Pu$_{2}$O$_{3}$ and how these properties change during the
reduction process from PuO$_{2}$ to $\alpha$-Pu$_{2}$O$_{3}$, it is
essential to first calculate the corresponding properties of
PuO$_{2}$. Table I shows our calculated lattice constant
\emph{a}$_{0}$ and bulk modulus \emph{B}$_{0}$ and its pressure
derivative \emph{B}$_{0}^{^{\prime}}$ within GGA+\emph{U} scheme
with different Hubbard \emph{U} parameters. We performed the
ferromagnetic (FM) and anti-ferromagnetic (AFM) coupling
calculations to decide which is the ground state according to the
total energies for each choice of the Hubbard \emph{U} value.
PuO$_{2}$ is known to be an insulator \cite{r15} and some scattered
experimental data proved its ground state to be an AFM phase
\cite{r16}. For bare GGA, i.e., \emph{U}=0 eV, our calculated result
predicts PuO$_{2}$ to be a FM metal, similar to the conclusion of
previous studies. This is contrary to experimental observation. By
increasing the amplitude of $U$, as shown in Fig. 2, the total
energy difference between the two phases decreases; the AFM phase
begins to be energetically preferred at $U\mathtt{\sim}$1.5 eV. We
find that the convergence to the correct AFM phase is very rapid.
Taking \emph{U}=3.0 eV, for example, the total energy (per formula
unit) for the AFM phase is lower than for the FM phase by a large
value of 0.4 eV. Note that in our calculation of PuO$_{2}$ in this
section, we have chosen two kinds of magnetic configurations. One is
that the Pu magnetic moments $\mu$ are confined along the $z$ axis
in a simple +$-$+$-$ alternation of spins, while the other is the
innerlayer alternation of spins. Our results show that the total
energies for the two configurations are degenerate. In the ideal
ionic limit, the formal charge for Pu ions in PuO$_{2}$ is +4,
corresponding to the formal population of \emph{f}$^{4}$ for
\emph{f} orbitals. Our direct calculated magnetic moments $\mu$ for
each plutonium ion is 4.16 (in unit of Bohr magneton) mainly
contributed by 4\emph{f} orbitals of 4.09, which is very close to
the population of \emph{f}$^{4}$.
\begin{figure}[ptb]
\includegraphics*[height=5.8cm,keepaspectratio]{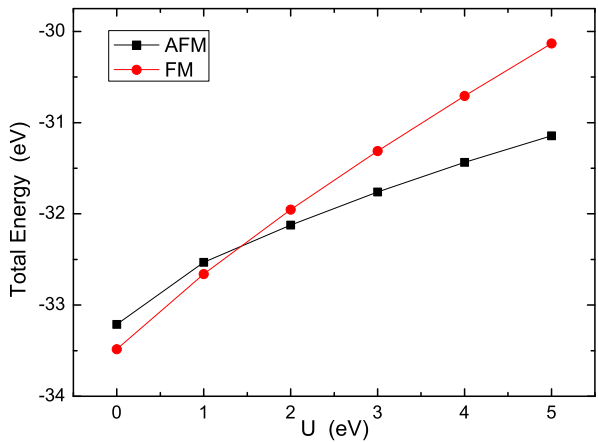}
\caption{The total energies for PuO$_{2}$ in FM and AFM phases with different Hubbard \emph{U} parameters.}%
\label{fig2:epsart}%
\end{figure}

\begin{table}[ptb]
\caption{Calculated and experimental lattice constants \emph{a}$_{0}$({\AA })
and bulk modules \emph{B}(GPa) for Pu$_{2}$O$_{3}$ and PuO$_{2}$ by the
GGA+\emph{U} scheme. }%
\begin{ruledtabular}
\begin{tabular}{lccccccc}
&\emph{U}(eV) & order&\emph{a}$_{0}$(\AA)   &  \emph{B}$_{0}$(GPa) & \emph{B}$_{0}^{'}$ \\
\hline
PuO$_{2}$      &4.0& FM&5.454&184&3.72\\
&4.0&AFM&5.477&184&3.72\\
&3.0& FM&5.438&183&4.72\\
&3.0&AFM&5.458&180&4.68\\
&0& FM&5.386&193&3.95\\
&0&AFM&5.398&187&3.67\\
&Expt.&&5.396&&\\
Pu$_{2}$O$_{3}$&0 & FM&10.91&133&5.52\\
&0 &AFM&10.92&123&4.25\\
&3.0& FM&11.14&119&4.05\\
&3.0&AFM&11.17&122&3.82\\
&4.0& FM&11.18&120&3.91\\
&4.0&AFM&11.20&128&3.31\\
&Expt.&&11.05&&\\
\end{tabular}
\end{ruledtabular}
\end{table}

The equilibrium lattice constant $a_{0}$ and bulk modulus $B_{0}$ and its
pressure derivative \emph{B}$_{0}^{^{\prime}}$ are obtained by fitting the
Murnaghan equation of state \cite{r18}. At \emph{U}=3.0 and 4.0 eV, the
present lattice constants \emph{a}$_{0}$ are 5.458 and 5.477 \AA \ for the AFM
phase, respectively, which are in good agreement with the experimental value
of 5.396 \AA \cite{r19}. As for the bulk modulus, our calculated results are
180 and 184 GPa with \emph{U}=3.0 and 4.0 eV, respectively, which agree well
with the recent refined experimental value of 179 GPa. However, in an early
experiment \cite{r10}, this value is largely overestimated to be 379 GPa.
Notice that the results of lattice constant and bulk modulus are 5.46
\AA \ and 220 GPa, respectively, which are obtained using
hybrid-density-functional calculations \cite{r20}. The physical insulating
behavior of PuO$_{2}$ with an experimentally comparable Mott gap of
$\mathtt{\sim}$1.5 eV has as well been obtained with Hubbard \emph{U} in a
range of 3-4 eV. Therefore, our GGA+\emph{U} results can provide a
satisfactory description of the atomic, mechanical, electronic, and magnetic
(FM or AFM) structures of PuO$_{2}$. This encourages us to investigate in the
following the ground-state properties of $\alpha$-Pu$_{2}$O$_{3}$ by the same
method, which turns to be remarkably effective as well. The present
calculation of PuO$_{2}$ is also necessary for theoretical access to
the\ PuO$_{2}\mathtt{\leftrightarrow}$Pu$_{2}$O$_{3}$($\alpha$) redox energy,
which will be discussed below.

\subsection{Atomic and electronic structure of $\alpha$-Pu$_{2}$O$_{3}$}

$\alpha$-Pu$_{2}$O$_{3}$ in body-centered cubic structure is the end product
during the reduction action process. The crystal structure of $\alpha$%
-Pu$_{2}$O$_{3}$ is well confirmed by X-ray powder diffraction measurement
\cite{r4,r5}. As mentioned above, after building the PuO$_{2}$ 2$\times
$2$\times$2 supercell and removing 25\% oxygen atoms in special sites,
$\alpha$-Pu$_{2}$O$_{3}$ can be obtained. Therefore, there are 32 plutonium
atoms and 48 oxygen atoms in the unit cell for $\alpha$-Pu$_{2}$O$_{3}$ (see
Fig. 1). The calculated structural and mechanical parameters within
GGA+\emph{U} scheme with different Hubbard \emph{U} parameters for $\alpha
$-Pu$_{2}$O$_{3}$ are also collected in table I. We also performed the
calculations in FM and AFM phases for $\alpha$-Pu$_{2}$O$_{3}$. For the AFM
calculations, we adopted two different spin arrangements of plutonium ions.
One spin arrangement is to confine $\mu$ along the \emph{z} axis in a simple
intra- or interlayer +$-$+$-$ alternation of spins, as what we dealt with
PuO$_{2}$ in the last section. Another model is to use the same spin
arrangement as in $\alpha$-Mn$_{2}$O$_{3}$. Based on the neutron diffraction
measurement, Regulski \emph{et al}. \cite{r22} proposed a appropriate
collinear AFM ordering, in which four basic magnetic cubic sublattices are
constructed and the summation of magnetic moments of eight ions are assumed to
be zero in each cube. We also adopt this AFM configuration as one additional
candidate to calculate the total energy of $\alpha$-Pu$_{2}$O$_{3}$ as a
function of $U$. The calculated lattice constant in the latter AFM ordering
are 10.92, 11.17, and 11.20 \AA \ for \emph{U}= 0, 3.0, and 4.0 eV, and the
corresponding bulk modulus are 123, 122, and 128 GPa, respectively. Note that
the corresponding experimental lattice constant is in a range of 11.03 to
11.07 \AA  \cite{r4,r5}. We find that similar to PuO$_{2}$, at Hubbard
\emph{U}=0 eV, the result obtained by bare GGA predicts $\alpha$-Pu$_{2}%
$O$_{3}$ to be a FM metal. Considering that traditional DFT approach within
the bare GGA scheme underestimates the strong on-site Coulomb repulsion of
plutonium 5\emph{f} electrons and can not accurately describe the localization
of 5\emph{f} electrons, therefore, we conclude this result obtained by bare
GGA is not correct although no magnetic susceptibility or neutron powder
diffraction data are available. Further calculations show that at Hubbard
\emph{U}=3.0 eV and 4.0 eV, the ground state of $\alpha$-Pu$_{2}$O$_{3}$ is an
AFM insulator (see below). Notice that our direct calculations show that the
$\alpha$-Pu$_{2}$O$_{3}$ in AFM phase is preferred to the latter spin
arrangement as mentioned above. The corresponding energy differences $\Delta
E$=\emph{E}$_{\mathrm{{AFM1}}}\mathtt{-}$\emph{E}$_{\mathrm{{AFM2}}}$ are 43
and 26 meV at \emph{U}=3.0 and 4.0 eV per unit cell, respectively. As for the
stability of AFM phase, the energy differences $\Delta E$=\emph{E}%
$_{\mathrm{{AFM2}}}\mathtt{-}$\emph{E}$_{\mathrm{{FM}}}$ are $\mathtt{-}$3.63
and $\mathtt{-}$5.80 eV at \emph{U}=3.0 and 4.0 eV per unit cell, respectively.

\subsection{Changes in the electronic properties during reduction process}

\begin{figure}[ptb]
\includegraphics*[height=7.8cm,keepaspectratio]{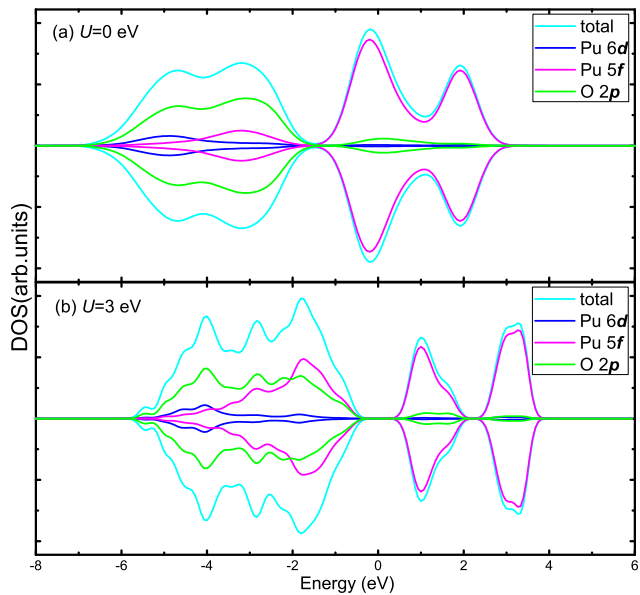}
\caption{The total DOS for AFM PuO$_{2}$ and projected orbital-resolved
partial DOS for Pu 6\emph{d}, Pu 5\emph{f}, and O 2\emph{p} at Hubbard (a)
\emph{U}=0 eV and (b)\emph{U}=3 eV, respectively. The Fermi level is set to
zero.}%
\label{fig2:epsart}%
\end{figure}

\begin{figure}[ptb]
\includegraphics*[height=7.8cm,keepaspectratio]{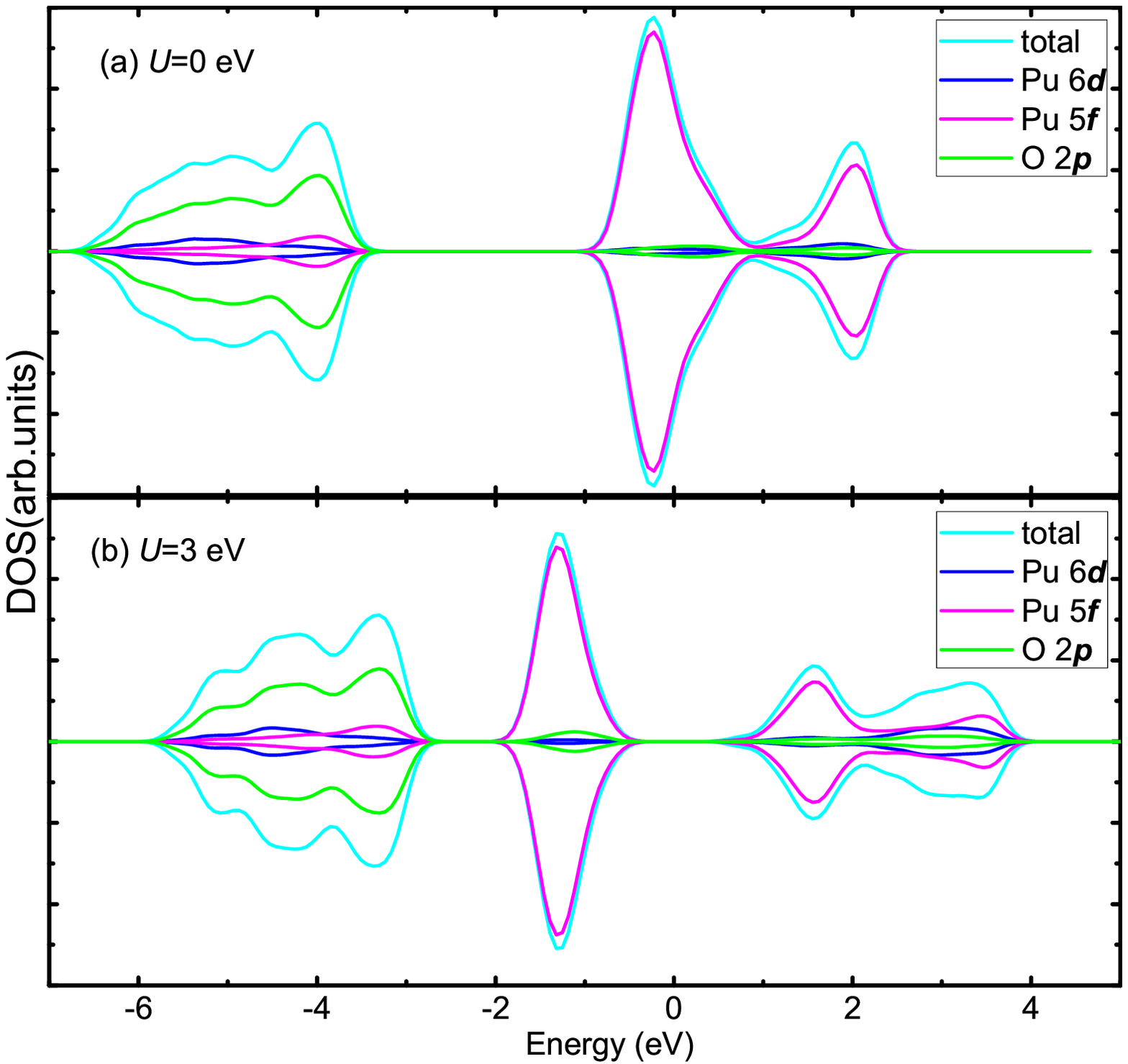}
\caption{The total DOS for AFM $\alpha$-Pu$_{2}$O$_{3}$ and projected
orbital-resolved partial DOS for Pu 6\emph{d}, Pu 5\emph{f}, and O 2\emph{p}
at Hubbard (a) \emph{U}=0 eV and (b)\emph{U}=3 eV, respectively. The Fermi
level is set to zero.}%
\label{fig3:epsart}%
\end{figure}

\begin{figure*}[ptb]
\includegraphics*[height=9.6cm,keepaspectratio]{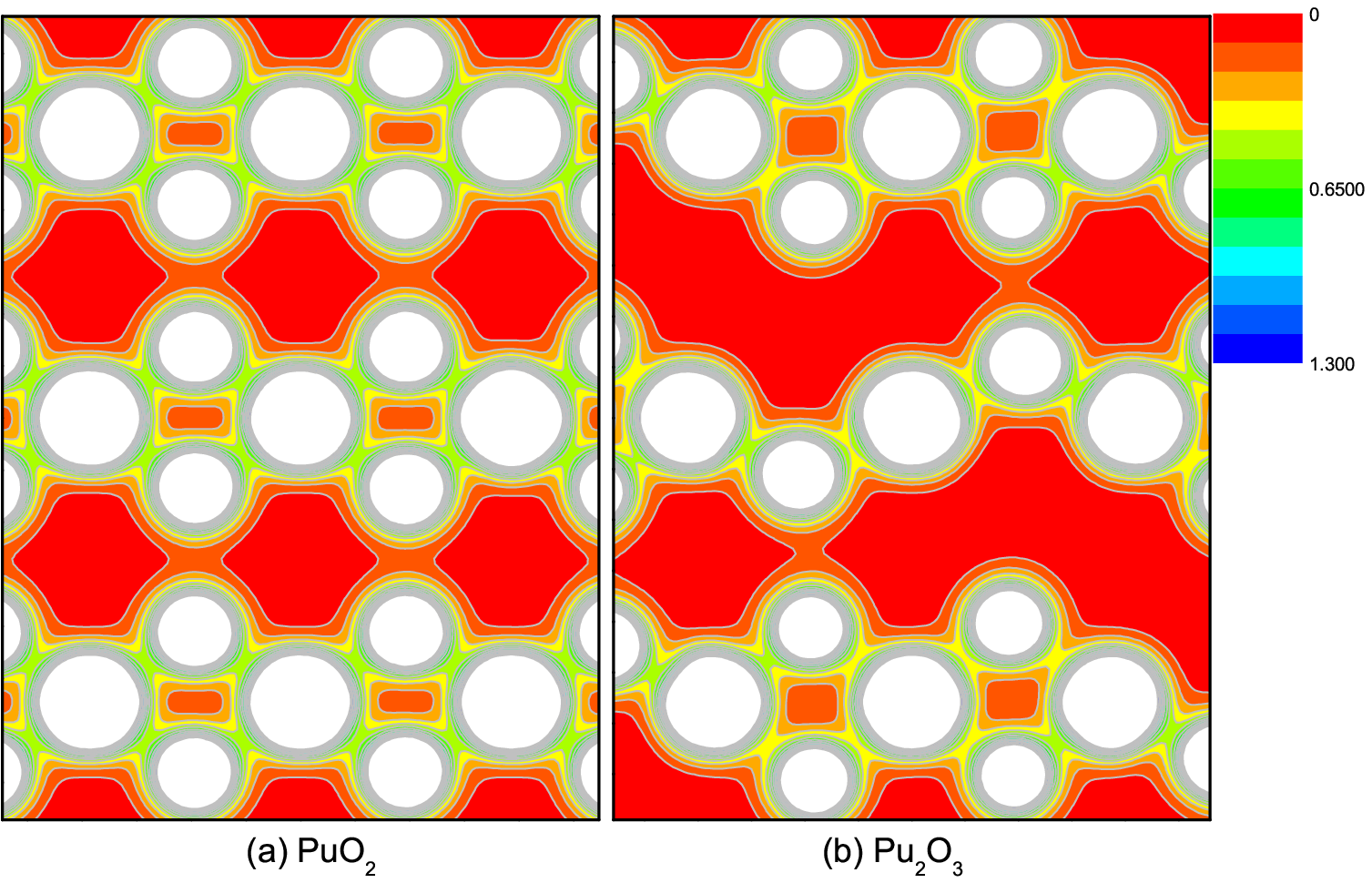}
\caption{Charge density in the (110) plane for (a) PuO$_{2}$ and (b) $\alpha
$-Pu$_{2}$O$_{3}$, respectively. The contour lines are plotted from 0.000 to
1.300 by the interval of 0.1083 e/\AA $^{3}$.}%
\label{fig4:epsart}%
\end{figure*}

In PuO$_{2}$ and $\alpha$-Pu$_{2}$O$_{3}$, the ideal oxidation states of Pu
ions are +4 and +3, respectively. This valence difference apparently shows
that the population of 5\emph{f} changes and in depth demonstrates that the
behavior of Pu 5\emph{f} is very complex. Directly, $\alpha$-Pu$_{2}$O$_{3}$
can be constructed from PuO$_{2}$ by removing 25\% oxygen atoms. After
removing one oxygen atom, two additional electrons are left behind. These two
electrons will occupy the lowest unoccupied states derived from Pu 5\emph{f}
orbitals. This leads to the fact that the population of \emph{f}$^{5}$ forms
in $\alpha$-Pu$_{2}$O$_{3}$, while \emph{f}$^{4}$ in PuO$_{2}$. This clearly
illustrates that Pu 5\emph{f} orbitals are more localized in $\alpha$-Pu$_{2}%
$O$_{3}$ compared with PuO$_{2}$. In the following, we systematically
investigated how the electronic structures characterized by 5\emph{f}
electrons change from PuO$_{2}$ to $\alpha$-Pu$_{2}$O$_{3}$. In order to make
comparison, the electronic structure properties of PuO$_{2}$ 2$\times$%
2$\times$2 supercell containing 96 atoms are also calculated. We will show in
detail how the volume of unit cell, density of states (DOS), charge
distribution, valence state, and magnetic order change. All the following
results are obtained by GGA+\emph{U} scheme with \emph{U}=3.0 eV except extra hints.

(I) The calculated lattice constants for PuO$_{2}$ and $\alpha$-Pu$_{2}$%
O$_{3}$ in their AFM phase are 5.458 and 11.17 \AA , leading to 7\% volume
expansion. This suggests that the 5\emph{f} electrons are more localized in
$\alpha$-Pu$_{2}$O$_{3}$. Due to the increasing localization, the interaction
between Pu ions and the cohesion of the crystal decrease, therefore the
lattice constant increases. The similar phenomenon has also been observed in
the reduction process from Ce$_{2}$O$_{3}$ to CeO$_{2}$ with a 10\% volume
change \cite{r6} also because of the increasing localization of the \emph{f} orbitals.

(II) The total DOS for PuO$_{2}$ (96 atoms) and
$\alpha$-Pu$_{2}$O$_{3}$ (80 atoms) within GGA+\emph{U} scheme at
different Hubbard \emph{U} are shown in Figs. 3 and 4, respectively.
In order to make a clear comparison, the orbital-resolved partial
DOS for Pu 6\emph{d}, Pu 5\emph{f}, and O 2\emph{p} orbitals are
also plotted. As showed in Fig. 3(a), at \emph{U}=0 eV, it is
clearly that the bare GGA scheme predicts PuO$_{2}$ to be a metal,
which is contrary to the experimentally established insulating
ground state \cite{r15}. However, in Fig. 3(b), at Hubbard
\emph{U}=3.0 eV, one can see that the present result correctly
predict PuO$_{2}$ to be an AFM insulator. As for
$\alpha$-Pu$_{2}$O$_{3}$, at Hubbard \emph{U}=0 eV showed in Fig.
4(a), the bare GGA approach predicts the FM metal ground state as we
expected, which is supposed to be not reasonable by considering the
necessity of the Hubbard \emph{U} correction. At Hubbard
\emph{U}=3.0 eV, the calculated DOS showed in Fig. 4(b) again
suggests $\alpha$-Pu$_{2}$O$_{3}$ to be in the insulating AFM ground
state. Concerning the satisfactory description of the electronic
structures for PuO$_{2}$ and the fact that the 5$f$ electrons in
$\alpha $-Pu$_{2}$O$_{3}$ are more localized than in PuO$_{2}$, we
predict $\alpha $-Pu$_{2}$O$_{3}$ to be an insulator in the AFM
state, although no experimental data can be obtained.

As for the projected orbital-resolved partial DOS showed in Fig.
3(b) for PuO$_{2}$, the occupied DOS is featured by two
well-resolved peaks. One near $\mathtt{-}$1.8 eV is dominated by Pu
5\emph{f} character, while another one near $\mathtt{-}$4.0 eV is
mostly O 2\emph{p}, which have observed in the recent photoemission
measurement \cite{r9}. One can see that Pu 6\emph{d}, Pu 5\emph{f},
and O 2\emph{p} states (especially between Pu 5\emph{f} and O
2\emph{p}) shows a significant hybridization covering from
$\mathtt{-}$5.7 to $\mathtt{-}$0.4 eV. This hybridization is also
responsible for the strong covalency of PuO$_{2}$. However, for
$\alpha$-Pu$_{2}$O$_{3}$, the Pu(5\emph{f})-O(2\emph{p})
hybridization is much smaller due to the fact that Pu 5\emph{f} and
O 2\emph{p} occupied states are well separated as showed in Fig.
4(b). Furthermore, the covalency in $\alpha$-Pu$_{2}$O$_{3}$ is
weaker than PuO$_{2}$, which can also be demonstrated by the charge
density distribution discussed below. We also note that in
$\alpha$-Pu$_{2}$O$_{3}$, the occupied 5\emph{f} peak around
$\mathtt{-}$1.3 eV is very narrow, indicating that the 5\emph{f}
orbitals are very localized, which is consistent with the conclusion
in above part (I).

(III) In order to obtain further understanding of the electronic
structure and bonding properties, the contours of the charge
densities in the (110) plane are also plotted for AFM PuO$_{2}$ and
AFM $\alpha$-Pu$_{2}$O$_{3}$ in Figs. 5(a) and 5(b), respectively.
Note that the contour lines are plotted from 0.000 to 1.300 by the
interval of 0.1083 e/\r{A}$^{3}$. For PuO$_{2}$ in Fig. 5(a), it is
evident that there are some closed contours existing between Pu and
O atoms. This suggests that the covalent bonding character exists in
PuO$_{2}$ as discussed in above part (II). Compared with the charge
density distribution for $\alpha$-Pu$_{2}$O$_{3}$ showed in Fig.
5(b), one can see that the charge density at most bridges connecting
Pu and O atoms are larger in PuO$_{2}$ than in
$\alpha$-Pu$_{2}$O$_{3}$, this demonstrates that the covalency is
much stronger in the former. Note that the distribution of charge
density around Pu atoms is nearly spherical, while the distribution
around O atoms is a little deformed towards their bonds. Therefore,
it is easy to decide their ionic radii according the minimum value
of charge density along the nearest Pu-O bond. Then we can obtain
the valence states for Pu ions discussed below.

(IV) For PuO$_{2}$ and $\alpha$-Pu$_{2}$O$_{3}$, there are 12.463 and 12.835
electrons around the plutonium ions with ionic radii of 1.287 and 1.319 \AA ,
therefore, the plutonium ions are presented as Pu$^{3.54+}$ and Pu$^{3.16+}$,
which are close to the corresponding ideal valence states +4 and +3 for
plutonium ions, respectively.

(V) As for the magnetic order, we notice for PuO$_{2}$
(2$\times$2$\times$2 supercell) the second spin arrangement model is
also more energetically
favored with the energy difference $\Delta E$=\emph{E}$_{\mathrm{{AFM1}}}%
$-\emph{E}$_{\mathrm{{AFM2}}}$ of 0.733 and 0.372 eV (per 2$\times$2$\times$2
supercell) at Hubbard \emph{U}=3.0 and 4.0 eV, respectively. As for $\alpha
$-Pu$_{2}$O$_{3}$, the corresponding energy differences $\Delta E$%
=\emph{E}$_{\mathrm{{AFM1}}}$-\emph{E}$_{\mathrm{{AFM2}}}$ are 43
and 26 meV at \emph{U}=3.0 and 4.0 eV per unit cell, respectively.
Our direct calculations show that the magnetic moments are about 4.2
and 5.0 $\mu_{B}$ per Pu ion in PuO$_{2}$ and
$\alpha$-Pu$_{2}$O$_{3}$, which are mainly contributed by 5\emph{f}
electrons of 4.1 and 5.0 $\mu_{B}$ and very close to the
corresponding populations of \emph{f}$^{4}$ and \emph{f}$^{5}$ for
plutonium ions, respectively. Notice that all the results of AFM
phase listed in Figs. 3, 4, and 5 are obtained by the second spin
arrangement model because it is energetically favored.

\subsection{Reduction reaction energy}

To further know the thermodynamic properties of the reduction process from
PuO$_{2}$ to $\alpha$-Pu$_{2}$O$_{3}$ via the reaction
\[
\mathrm{{PuO_{2}(2\times2\times2\ supercell)-8O_{2}=Pu_{2}O_{3}}(\alpha),}%
\]
we have calculated the reaction energy $\Delta$\emph{E}=\emph{E}%
$_{\mathrm{{Pu_{2}O_{3}}}}$-\emph{E}$_{\mathrm{{PuO_{2}}}}$+8\emph{E}%
$_{\mathrm{{O_{2}}}}$ at Hubbard \emph{U}=3.0 eV. Note that \emph{E}%
$_{\mathrm{{Pu_{2}O_{3}}}}$, \emph{E}$_{\mathrm{{PuO_{2}}}}$, and
\emph{E}$_{\mathrm{{O_{2}}}}$ are the total energies for Pu$_{2}$O$_{3}$ (unit
cell), PuO$_{2}$ (2$\times$2$\times$2 supercell), and (spin-polarized) oxygen
molecule, respectively. Considering the overestimation of the binding energy
for O$_{2}$ introduced by DFT, we use the experimental cohesive energy value
of 5.21 eV for O$_{2}$ \cite{r23}. Finally, the reaction energy is 55.04 eV
(per $\alpha$-Pu$_{2}$O$_{3}$ unit cell). This indicates that the reaction is
endothermic, which is consistent with the fact that $\alpha$-Pu$_{2}$O$_{3}$
can be experimentally obtained by partial reduction of PuO$_{2}$ at high
temperature \cite{r5}.

\section{SUMMARY}

In summary, we have studied the structural, electronic, and
thermodynamics properties of PuO$_{2}$ and $\alpha$-Pu$_{2}$O$_{3}$
in their antiferromagnetic insulator states. We find that after
partial oxygen atoms removed from PuO$_{2}$, the left electrons are
localized into the Pu 5\emph{f} orbitals. As a consequence, several
physical properties have changed correspondingly listed in
following. (1) In PuO$_{2}$, the ideal valence state for Pu is +4,
while in $\alpha$-Pu$_{2}$O$_{3}$, the valence state is +3. Our
charge density integration around Pu ion sphere also confirms this
conclusion. (2) The magnetic moments per Pu ion increase from 4
$\mu_{B}$ for PuO$_{2}$ to 5 $\mu_{B}$ for $\alpha$-Pu$_{2}$O$_{3}$,
which correspond to the populations of \emph{f}$^{4}$ and
\emph{f}$^{5}$, respectively. (3) Due to the more localization of
5\emph{f} electrons, which do not contribute to the chemical
bonding, thus the cohesion of $\alpha$-Pu$_{2}$O$_{3}$ decreases. As
a result, the volume expands about 7\% according to our calculation.
(4) Because of more localization of 5\emph{f} orbitals, the occupied
Pu 5\emph{f} and O 2\emph{p} orbitals are well separated, the
hybridization between them decreases. Finally the covalency of the
Pu-O bond in $\alpha$-Pu$_{2}$O$_{3}$ is weakened. The calculated
DOS and charge density distribution also demonstrate this.

\begin{acknowledgments}
This work was supported by the Foundations for Development of
Science and Technology of China Academy of Engineering Physics.
\end{acknowledgments}


\begin{thebibliography}{99}                                                                                               %


\bibitem {r1}S. Y. Savrasov and G. Kotliar, Phys. Rev. Lett. \textbf{84}, 3670(2000).

\bibitem {r2}K. T. Moore and G. van der Laan, Rev. Mod. Phys. 81, 235 (2009).

\bibitem {r3}J. M. Haschke, T. H. Allen, and L. A. Morales, Los Alamos Sci.
26, 253 (2000).

\bibitem {r4}W. H. Zachariasen, Metallurgical Laboratory Report, CK-1367, 1944.

\bibitem {r5}IAEA technical reports series, No. 79 (1967).

\bibitem {r6}N.V. Skorodumova, S. I. Simak, B. I. Lundqvist, I. A. Abrikosov,
and B. Johansson, Phys. Rev. Lett. \textbf{89}, 166601 (2002).

\bibitem {r7}S. L. Dudarev, G. A. Botton, S. Y. Savrasov, C. J. Humphreys, and
A. P. Sutton, Phys. Rev. B \textbf{57}, 1505 (1998).

\bibitem {r8}B. Sun, P. Zhang, and X.-G. Zhao, J. Chem. Phys. 128, 084705 (2008).

\bibitem {r9}M. Butterfield, T. Durakiewicz, E. Guziewicz, J. Joyce, A. Arko,
K. Graham, D. Moore, and L. Morales, Surf. Sci. \textbf{571}, 74 (2004).

\bibitem {Jom}G\'{e}rald. Jomard, B. Amadon, Francois Bottin, and M.
Torrent, Phys. Rev. B \textbf{78}, 075125 (2008).

\bibitem{Andersson}D. A. Andersson, J. Lezama, B. P. Uberuaga, C. Deo,
and S. D. Conradson, Phys. Rev. B \textbf{79}, 024110 (2009).

\bibitem{Petit}L. Petit, A. Svane, Z. Szotek, W. M. Temmerman, and G. M. Stocks,
arXiv:0908.1806v1 [cond-mat.str-el]
\bibitem {r90}M. Idiri, T. Le Bihan, S. Heathman, and J. Rebizant, Phys. Rev.
B \textbf{70}, 014113 (2004).

\bibitem {r10}J.-P. Dancausse, E. Gering, S. Heathman, and U. Benedict, High
Pressure Research \textbf{2}, 381 (1990).

\bibitem {r11}G. Kresse and J. Hafner, Phys. Rev. B \textbf{48}, 13115 (1993).

\bibitem {r12}Y. Wang and J.P. Perdew, Phys. Rev. B \textbf{44}, 13298 (1991).

\bibitem {r13}G. Kresse, D. Joubert, Phys. Rev. B \textbf{59}, 1758 (1999).

\bibitem {r15}C. E. McNeilly, J. Nucl. Mater. 11, 53 (1964).

\bibitem {r16}P. Santini, R. Lemanski, and P. Erdos, Adv. Phys. \textbf{48},
537 (1999); M. Colarieti-Tosti, O. Eriksson, L. Nordstrom, J. Wills, and M. S.
S. Brooks, Phys. Rev. B \textbf{65}, 195102 (2002).

\bibitem {r18}F. Birch, Phys. Rev. \textbf{71}, 809 (1947).

\bibitem {r19}R. G. Haire and J. M. Haschke, MRS Bull. \textbf{26}, 689 (2001).

\bibitem {r20}I. D. Prodan, G. E. Scuseria, and R. L. Martin, Phys. Rev. B
\textbf{73}, 045104 (2006).

\bibitem {r21}J. Cable, M. Wilkinson, E. Woolan, W. Koehler, Phys. Prog. Rep.
\textbf{43} ORNL-2302 (1957).

\bibitem {r22}M. Regulski, R. Przenioslo, I. Sosnowska, D. Hohlwein, R.
Schneider, J. Alloy. Compoun. \textbf{362}, 236 (2004).

\bibitem {r23}K. Huber, American Institue of Physics Handbook, edited by D. E.
Gray(McGraw-Hill, New York, 1972).
\end{thebibliography}
\end{document}